\documentclass[aps,showkeys,showpacs,amsmath,amssymb]{revtex4}
\usepackage{color}
\usepackage{graphicx}
\input{Qcircuit}
\begin{document}

\title[Four qubit two magnon states] {Multipartite entangled magnon states as quantum communication channels}

\author{Sriram Prasath E.$^{1}$, Sreraman Muralidharan$^{2}$, Chiranjib Mitra$^{3}$ and Prasanta K. Panigrahi$^{3,4}$ }

\address{$^{1}$Maulana Azad National Institute of Technology, Bhopal - 462051, India}
\address{$^{2}$Loyola College, Chennai - 600034, India}
\address{$^{3}$Indian Institute of Science Education and Research - Kolkata, Mohanpur, BCKV Campus Main Office, Mohanpur - 741252, India}
\address{$^{4}$ Physical Research Laboratory, Navrangpura, Ahmedabad-380 009, India}
\email{prasanta@prl.res.in}
\begin{abstract}
We explicate conditions under which, the two magnon state becomes
highly entangled and is useful for several quantum communication protocols. This state,
which is experimentally realizable in quantum dots using Heisenberg
exchange interaction, is found to be suitable for carrying out
deterministic teleportation of an arbitrary two qubit composite
system. Further, conditions for which the channel capacity reaches
``Holevo bound", allowing  maximal amount of classical information
to be encoded, are derived. Later, an explicit protocol for the splitting and sharing of a two qubit entangled
state among two parties, using this state as an entangled resource, is demonstrated.

\end{abstract}

\maketitle
Entanglement, an entirely quantum mechanical phenomenon, has been a
subject of intense research, since the days of Schr\"{o}dinger
\cite{ZPhys} and Einstein \cite{EPR Paper}. In recent times, it has
received renewed attention with the advent of quantum information
processing \cite{Nc}. The amalgamation of the principles of quantum
entanglement and information theory has led to the possibility of
carrying out tasks, which would have been otherwise impossible in
the classical world \cite{tele, dense}. The fact that some of the finite dimensional
spin states naturally occur in physical systems and these can be
experimentally realized makes them the states of choice for
explicating quantum protocols \cite{Bose}. The search for physically occurring
entangled systems, which can be used for carrying out these quantum
tasks, is of interest to both theorists and experimentalists.

A natural system attracting considerable interest is that of spin
chains, where the Heisenberg exchange interaction can create desired
entangled states \cite{cluster1}. Quantum dots have also shown considerable promise
in the realization of these states, which can be manipulated by
varying the voltage applied through gate electrodes between adjacent
quantum dots \cite{qdot}. Interestingly, entanglement has been found to be
retained even up to certain non-vanishing temperature. Entangled states e.g.,
Bell states $|\phi_{\pm}\rangle = \frac{1}{\sqrt{2}},
(|01\rangle\pm|10\rangle)$ multipartite cluster states \cite{cluster1, cluster2}:
\begin{equation}
|C_N\rangle = \frac{1}{2^{N/2}} \otimes_{a=1}^{N} (|0\rangle_a \sigma_z ^{a+1} + |1\rangle_a)
\end{equation}
have found application in carrying out various quantum
tasks like error correction \cite{cluster2} and state
sharing \cite{sree2}.

In a spin chain, low lying magnon excitations of the ground state
can show robust entanglement properties. In fact, the well known
Bell states and $W$ states of the form \cite{Zheng, Pati, Rao1}
\begin{equation}
|W'\rangle = (\alpha|100\rangle+\beta|010\rangle+\gamma|001\rangle),
\label{W-state}
\end{equation}
are natural one magnon states, above the ferromagnetic ground states
formed of two and three particles respectively. Although Bell states
can be used for achieving perfect teleportation of, $|\psi_1\rangle =  \alpha|0\rangle+\beta|1\rangle$,
$(\alpha,\beta \in C, |\alpha|^2 + |\beta|^2 = 1)$ the symmetric W states cannot be used. However, $|W'\rangle$ can be
used for the perfect teleportation of $|\psi_1\rangle$ in the case
where $|\alpha|^2 + |\beta|^2 = |\gamma|^2$. Hence, the characterization of
 different multipartite magnon states need careful investigation for
diverse quantum tasks.
In this paper, we systematically investigate the four qubit, two magnon
state for its ability to carry out various quantum tasks like
teleportation, state sharing and dense coding. We start with the non
trivial task of teleporting $|\psi_2\rangle = \sum_{0}^{1}
\alpha_{i_{1}i_{2}}|i_{1}i_{2}\rangle$ $(\alpha_{i_{1}i_{2}} \in C, \sum|\alpha_{i_{1}i_{2}}|^2 = 1)$, for which one needs two ebits of
entanglement between the parties. The conditions for achieving the
same are also discussed. It is worth mentioning that in case of four particles, only one of the
nine distinct classes, characterized by LOCC \cite{9ways}, can be used for the teleportation of an arbitrary two qubit state.

The most general two magnon four qubit state is given by,
\begin{eqnarray}
|4;2\rangle = \sum_{i=0}^{1}
(W_{ii\tilde{i}}|ii\tilde{i}\tilde{i}\rangle +
W_{i\tilde{i}i}|i\tilde{i}i\tilde{i}\rangle +
W_{\tilde{i}ii}|i\tilde{i}\tilde{i}i\rangle)_{1234} \label{Gen_W_state}
\end{eqnarray}
where $\tilde{i}$ is the negation of $i$.

Here the first index on the left hand side denotes the number of
qubits in the channel and the second one the number of magnons. From
an experimental point of view, spins of electrons in quantum dots
are proposed to be good building blocks for obtaining the above type of states
\cite{qdot, petta1, loss}. With a high level of control over the number of
electrons \cite{ashoori}, and the applied external fields, desirable
initial states can be achieved. The exchange interaction between any
two electronic spins can be controlled by the applied gate voltages
\cite{petta1, petta2}. Specific multi-qubit entangled states can be
realized by switching on the exchange interaction, generated by the
following Hamiltonian:
\begin{equation}
H = J S_A . S_B + J S_B . S_C + J \Delta S_A . S_C,
\label{Heisenberg}
\end{equation}
where J is the strength of interaction. States having definite spin,
e.g., generalised W states can be created using different values of
the parameter $\Delta$, at different points in time due to exchange
interaction induced dynamics. For example,  $|W'\rangle$ can be
obtained when $\Delta=1$ and at time $t = \frac{2}{3J} cos ^{-1}
\left(\frac{1}{8}\right)$ \cite{Rao1}. $N$-qubit $W$ states can, in
general, be formed by varying the exchange interaction using the
gate voltage in an array of quantum dots and also allowing that
variation to happen on a fixed time scale. The amplitudes $W_{ijk}$'s in
Eq. (\ref{Gen_W_state}) can be varied by suitably choosing the time
scales over which, the gate electrodes can be switched on and off. The generalized $W$ states can also be
generated from other interactions which conserve the $\hat{z}$
component of the total spin of qubits.
The exchange interaction induced quantum gates, for example,
${SWAP}$ and $CNOT$ gates have been implemented at picosecond time
scales \cite{qdot}.

This procedure can be generalized to $N$-magnon entangled channels.
For specificity, we concentrate here on the two magnon state and
explore its utility for various quantum tasks, starting from teleportation in the following section.

\section{Teleportation of $|\psi_2\rangle$}

In this protocol, Alice possesses qubits 1 and 3, and
Bob qubits 2 and 4 of $|4;2\rangle$ respectively. Alice performs a
joint four qubit von-Neumann type measurement on $|\psi_2\rangle$ and on her qubits. Let the qubits in $|\psi_2\rangle$ be represented by subscripts p and q and the four qubits from $|4;2\rangle$ be represented by subscripts 1, 2, 3 and 4.
The combined state of $|\psi_2\rangle_{pq}$ and $|4;2\rangle_{1234}$ can be written as

\begin{eqnarray}
\sum_{i = 1}^{16} (M_i \otimes N_i \otimes I \otimes I) |4;2\rangle'_{pq13})  (N_i \otimes M_{(i+8)mod(16)} |\psi_2\rangle_{24}),
\end{eqnarray}
where $N_i, \ M_i $ are given in table 1 and
\begin{eqnarray}
|4;2\rangle' = \sum (W_{\tilde{i}\tilde{i}i}|ii\tilde{i}\tilde{i}\rangle + W_{i\tilde{i}i}|i\tilde{i}i\tilde{i}\rangle + W_{\tilde{i}ii}|i\tilde{i}\tilde{i}i\rangle).
\end{eqnarray}

 The set of operators $ M_i \otimes N_i\otimes I \otimes I$, for values of $i$ from 1 to 16, when operated on $|4;2\rangle'$ generate the measurement basis for Alice. Since these sixteen states need to be distinguishable, they have to be mutually orthogonal.  The orthogonality condition results in the following  relations,
\begin{eqnarray}
W_{011}W_{110}^*+W_{001}W_{100}^*=0,\\ \nonumber
 |W_{101}|^2 = |W_{110}|^2+ |W_{100}|^2 =  |W_{011}|^2 + |W_{010}|^2=|W_{001}|^2. \\ \nonumber
\end{eqnarray}
The resulting state $|4;2\rangle$ fulfilling these conditions possesses two ebits of entanglement between the subsytems $(1,3)$
and $(2,4)$ respectively. Alice, now conveys the outcome of her measurement to Bob via four classical bits. After receiving the
outcome of Alice's measurement, Bob can perform an appropriate unitary operation $ (N_i \otimes M_{(i+8)mod(16)})^{-1}$
on his qubits and obtain $|\psi_2\rangle$.
\begin{table}
\caption{\label{tab1}Required operators for generating Alice's measurement basis.}
\begin{tabular}{|c|c|c|}
\hline
\textbf{$i$} &\textbf{operator $M_i$} &\textbf{ operator $N_i$}\\
\hline
$1$&$\sigma_0$&$\sigma_0$\\
$2$&$\sigma_0$&$\sigma_3$\\
$3$&$\sigma_3$&$\sigma_0$\\
$4$&$\sigma_3$&$\sigma_3$\\

$5$&$\sigma_1$&$\sigma_0$\\
$6$&$\sigma_1$&$\sigma_3$\\
$7$&$\sigma_2$&$\sigma_3$\\
$8$&$\sigma_2$&$\sigma_0$\\

$9$&$\sigma_0$&$\sigma_1$\\
$10$&$\sigma_0$&$\sigma_2$\\
$11$&$\sigma_3$&$\sigma_1$\\
$12$&$\sigma_3$&$\sigma_2$\\

$13$&$\sigma_1$&$\sigma_1$\\
$14$&$\sigma_2$&$\sigma_1$\\
$15$&$\sigma_1$&$\sigma_2$\\
$16$&$\sigma_2$&$\sigma_2$\\

\hline
\end{tabular}
\end{table}

From the earlier conditions, it can be seen that the members of Alice's measurement basis are mutually orthogonal
to each other. Hence these states can be perfectly distinguished making the present scheme deterministic.
It is worth mentioning that teleportation can also be
successfully implemented if we redistribute the qubits between Alice and Bob.
We shall now turn our attention towards the usefulness
of $|4;2\rangle$ for dense coding.

\section{Dense coding}
The general condition for a  given $2N$ qubit entangled channel $|\xi_{AB}\rangle$ to be used
for sending $2N$ classical bits by sending $N$ quantum bits is that there
has to be $N$ ebits of entanglement between $A$ and $B$.
In the dense coding scenario, we let Alice possess qubits 1 and 3, while Bob is
left with 2 and 4. Alice applies a set of unitary operators from
$(\sigma_0, \sigma_1, i\sigma_2, \sigma_3)$ and encodes her classical
information in operators and sends her qubits to Bob. The sixteen states obtained by Bob are
\begin{equation}
( M_{i} \otimes I \otimes  N_{i} \otimes I )|4;2\rangle.
\end{equation}

The amount of classical bits that can be encoded
into a given quantum state $\rho^{AB}$, shared by Alice and Bob, is given by \cite{Bruss},
\begin{equation}
X(\rho^{AB})=\mathrm{log_{2}}d_{A}+S(\rho^{B})-S(\rho^{AB})
\end{equation}
Here, $d_A$ refers to the dimension of the Alice's system. The maximal amount of
classical information that could be encoded in a four qubit quantum state is four.
When $X(\rho^{AB})$ reaches the maximum value, the protocol is said to reach
the ``Holevo bound". $|4;2\rangle$ reaches the ``Holevo bound" when all the sixteen states obtained by Bob are distinguishable i.e., all of them are orthogonal. Subjecting the states to be orthogonal, we have the following conditions:
\begin{eqnarray}
|W_{101}|^2=|W_{010}|^2=|W_{001}|^2=|W_{110}|^2+|W_{011}|^2 \\ \nonumber
\textrm{and} \, W_{100}=0.
\end{eqnarray}

\begin{figure}[h]
	\caption{Circuit diagram to experimentally generate $|4;2\rangle$, satisfying relations for dense coding.}
	\label{fig:Circuit}
	\leavevmode
\centering	
\Qcircuit @C=1em @R=.7em {
\lstick{\ket{0}} &\qw  &\qw     &\qw &\qw    &\qw &\targ        &\targ &\qw  &\ctrl{4}&\qw &\qw  &\qw&\qw   &\targ  &\qw&\qw&\qw &\qw&\ctrl{4} &\qw &\qw\qwx[3]\\
\lstick{\ket{0}} &\qw  &\qw     &\qw &\qw    &\targ&\qw       &\qw &\targ  &\qw&\ctrl{3}&\qw &\qw &\ctrl{3}  &\qw &\qw &\gate{\sigma_X}&\targ&\qw &\qw &\qw&\qw\\
\lstick{\ket{0}} &\qw  &\gate{H}&\qw &\ctrl{2}&\qw&\qw        &\qw &\ctrl{-1} &\qw &\qw&\qw  &\ctrl{2} &\qw &\qw  &\qw &\qw &\qw &\ctrl{2}&\qw &\qw&\qw\\
\lstick{\ket{0}} &\qw  &\gate{H}&\ctrl{1}&\qw &\qw &\qw       &\ctrl{-3}&\qw &\qw  &\qw &\qw  &\qw&\qw &\targ&\gate{H}&\qw&\ctrl{-2} &\qw &\qw&\qw&\qw\\
\lstick{\ket{0}_a}&\qw&\qw &\targ &\targ   &\ctrl{-3} &\ctrl{-4}&\ctrl{-1}&\ctrl{-2}&\targ&\targ  &\gate{\sigma_X}&\targ&\targ&\ctrl{-4}&\ctrl{-1}&\ctrl{-3}&\ctrl{-1} &\targ &\targ &\qw &\rstick{\ket{0}_a}\\
}
\end{figure}
An experimental procedure, to generate $|4;2\rangle$ satisfying these relations, is shown as a quantum circuit in Fig.\ref{fig:Circuit}.
\\
 In order to obtain the classical information encoded by Alice, Bob performs a joint von-Neumann type measurement. Since, all these states are orthogonal to each other,
they can be distinguished perfectly. Thus, Alice sends four classical bits
by sending only two quantum bits making the superdense coding
capacity reach the ``Holevo bound".

\section{Quantum Information splitting of an entangled state}
Entanglement can be used for splitting and sharing of
both classical and quantum information.
Sharing of quantum information between a group
of parties such that none of them can reconstruct the
unknown information by operating on their own qubits is
referred to as ``Quantum information splitting ".

Quantum information splitting of a single qubit state
$|\psi_1\rangle$ was first achieved using a three qubit GHZ state,
$|GHZ\rangle = \frac{1}{\sqrt{2}} (|000\rangle \pm |111\rangle)$, as
a shared entangled resource \cite{hillery, Bandy}. In this scheme the three parties involved,
Alice, Bob and Charlie, possess one qubit each. Alice has an unknown
qubit $|\psi_1\rangle$, that she wants Bob and Charlie to share. To achieve
this purpose, Alice performs a Bell measurement on her qubits and conveys
its outcome to Charlie via two classical bits. Now, the unknown qubit
is locked between Bob and Charlie such that none of them can obtain it completely
by operating on their own qubits. Now, if Bob can perform a measurement in a suitable basis and can convey his outcome
to Charlie via one classical bit, then Charlie can reconstruct $|\psi_1\rangle$ by operating
on his qubits. This will complete the QIS protocol of $|\psi_1\rangle$.

The feasibility of realizing QIS experimentally was discussed by making use
of a pseudo GHZ state \cite{tittel}. Experimental schemes which used single photon sources
to split $|\psi_1\rangle$ have been demonstrated \cite{schmid}. Recently, attention has turned
towards the usage of genuine multipartite entangled channels which can be
realized experimentally \cite{sree3, Sayan}.
Further, it was proved that one can devise $(N-2n)$ protocols for the QIS of
an arbitrary $n$ qubit state using a genuinely entangled $N$ qubit state
as an entangled channel among two parties, in the case where the two parties need
not meet \cite{Srikant}. According to this theorem, one cannot use a four qubit
channel for QIS of an arbitrary two qubit state. Interestingly, here
we show that $|4;2\rangle$ can be used for the QIS of an entangled state
of the type $A|00\rangle+B|11\rangle$, if we devise an unconventional QIS protocol.
The normalized $|4;2\rangle$ state fulfilling the following
conditions,
\begin{eqnarray}
|W_{110}|^2 = |W_{100}|^2 + |W_{010}|^2, \\
|W_{001}|^2=|W_{101}|^2 +|W_{011}|^2,\\
W_{110}W_{001}^*=W_{100}W_{011}^*+W_{010}W_{101}^{*}=0,
\end{eqnarray}
can be used for QIS of $A|00\rangle+B|11\rangle$.
In this scenario, Alice has the first qubit, Bob has the second
qubit and Charlie is left with the last two qubits. Alice performs a
joint measurement on her qubits and conveys its outcome to Charlie
via two classical bits. Bob performs a measurement on his qubit
in $(|0\rangle, |1\rangle)$ basis and conveys his outcome to Charlie
via one classical bit. The outcome of Alice's and Bob's measurement
and the state obtained by Charlie are shown in table 2.
\begin{small}
\begin{table}[h]
\caption{\label{tab4} QIS of an entangled state using $|4;2\rangle$. Normalization factors have been removed for convenience.}
\begin{tabular}{|c|c|c|}
\hline
\textbf{Alice's measurement} & \textbf{Bob's measurement}&\textbf{State obtained by Charlie}\\
\hline
$\frac{|001\rangle+ |110\rangle}{\sqrt{2}} $&$|1\rangle$&$ AW_{110}|00\rangle+B(W_{100}|10\rangle+W_{010}|01\rangle)$ \\
$$&$|0\rangle$&$ A(W_{101}|10\rangle+W_{011}|01\rangle)+BW_{001}|11\rangle$\\
\hline
$\frac{|001\rangle- |110\rangle}{\sqrt{2}}$&$|1\rangle$&$ AW_{110}|00\rangle-B(W_{100}|10\rangle+W_{010}|01\rangle)$\\
$$&$|0\rangle$&$ A(W_{101}|10\rangle-W_{011}|01\rangle)+BW_{001}|11\rangle$\\
\hline
$\frac{|111\rangle+ |000\rangle}{\sqrt{2}} $&$|1\rangle$&$ BW_{110}|00\rangle+A(W_{100}|10\rangle+W_{010}|01\rangle)$\\
$ $&$|0\rangle$&$ B(W_{101}|10\rangle+W_{011}|01\rangle)+AW_{001}|11\rangle$\\
\hline
$\frac{|111\rangle- |000\rangle}{\sqrt{2}} $&$|1\rangle$&$ BW_{110}|00\rangle-A(W_{100}|10\rangle+W_{010}|01\rangle)$\\
$$&$|0\rangle$&$ B(W_{101}|10\rangle+W_{011}|01\rangle)-AW_{001}|11\rangle$\\
\hline
\end{tabular}
\end{table}
\end{small}

From table 2, it can be observed that Charlie still does not possess a state, which can
be converted to $A|00\rangle+B|11\rangle$ through local operations on his individual qubits.

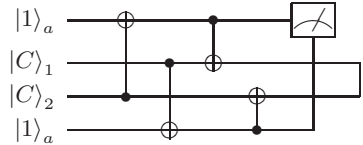
\begin{figure}[h]
	\caption{Circuit diagram to retrieve the desired state if Bob measures $|0\rangle$.}
	\label{fig:CircuitDiagram}
	\leavevmode
\centering	
\Qcircuit @C=1em @R=.7em {
\lstick{\ket{1}_a}  &\qw &\targ    &\qw     &\ctrl{1} &\qw      &\meter\\
\lstick{\ket{C}_1}  &\qw &\qw      &\ctrl{2}&\targ    &\qw      &\qw&\qw\\
\lstick{\ket{C}_2} &\qw  &\ctrl{-2}&\qw     &\qw      &\targ    &\qw&\qw\qwx\\
\lstick{\ket{1}_a} &\qw  &\qw      &\targ   &\qw      &\ctrl{-1}&\qw\qwx[-3]\\
}
\end{figure}

Hence, Charlie has to carry out further operations at his end to obtain $A|00\rangle+B|11\rangle$.
If Bob measures $|0\rangle$ then, Charlie needs to carry out a two-step process to obtain the desired
state, on his qubits $|C\rangle_1$ and $|C\rangle_2$. First, Charlie attaches two ancilla bits to his two qubit
state. Subsequently, he needs to carry out required operations, as
shown in Fig. \ref{fig:CircuitDiagram} and given in table
\ref{tab5}. The disentangling operation performed by Charlie using two ancilla qubits is
unconventional, after which Charlie ends up with a four qubit
state.  He then needs to perform a two qubit measurement with a
basis, which depends on the input magnon state $|4;2\rangle$, as described in the table 3. The measurement is done over the ancilla qubits and the final state is obtained on the qubits $|C\rangle_1$ and $|C\rangle_2$.

\begin{table}[h]
\caption{\label{tab5}Charlie's disentangling operations.}
\begin{tabular}{|c|c|c|}
\hline \bf{Bob's measurement outcome}& \bf{Charlie's measurement basis}& \bf{final state}\\ \hline

$|1\rangle$ &$W_{110}|00\rangle + W_{100}|01\rangle + W_{010}|10\rangle$& $A|00\rangle+B|11\rangle$\\

$$ & $W_{110}|00\rangle -W_{100}|01\rangle - W_{010}|10\rangle$&$A|00\rangle-B|11\rangle$\\
 \hline
 $|0\rangle$&$W_{101}|10\rangle + W_{011}|01\rangle + W_{001}|00\rangle$&$B|00\rangle+A|11\rangle$\\
 $$&$W_{101}|10\rangle +W_{011}|01\rangle - W_{001}|00\rangle$&$B|00\rangle-A|11\rangle$\\
 \hline
\end{tabular}
\end{table}
Here, $|C\rangle_1$ and $|C\rangle_2$ refer to Charlie's first and
second qubits respectively, and $|1\rangle_{a}$ refers to the ancilla
used by Charlie. If Bob measures $|1\rangle$, then the same circuit
diagram is implemented using $|0\rangle_a$ as the ancilla bit.
This completes the protocol for the QIS of an entangled state
using $|4;2\rangle$ as an entangled channel.

\section{Conclusion}
We have investigated the conditions under which the four qubit two magnon
state  $|4;2\rangle$ gets maximally entangled and illustrated its usefulness for several quantum communication protocols.
After establishing general conditions for which the state
$|4;2\rangle$ can be used for the deterministic quantum
teleportation of a two qubit composite system, we showed that this state can
also be used for required splitting up
of a two qubit entangled state among two parties. Further, conditions are established for which the
channel capacity of this state reaches the ``Holevo bound", allowing
four classical bits to be transmitted with just two qubits. Since
teleportation of a two qubit composite system has been
experimentally demonstrated using Bell pairs and $|4;2\rangle$ can
be realized using Heisenberg exchange interactions, all the
presented schemes are experimentally feasible in either spin chains or in an array of coupled quantum dots. These protocols can be extended to other quantum channels
in the future. Further, we wish to investigate the usefulness of magnon
states for quantum secure direct communication. We also intend to
find out the usefulness of higher particle magnon states for quantum
communication protocols. It has been shown, that in the presence of decoherence, the fidelity of
teleportation in a channel depends upon the coupling of the quibits to a common bath \cite{durga-mitra}.
In future, we intend to investigate the effect of decoherence in the present system. We also plan to carry out these protocols by performing a non
destructive discrimination scheme on multiple qubits instead of joint
multiqubit entangled measurements.

\end{document}